%% file: Instrument_Paper10.tex
\documentclass[aps,amssymb,amsmath,pra,twocolumn,showpacs,superscriptaddress]{revtex4-1}

\usepackage{graphicx}% Include figure files
\usepackage{dcolumn}% Align table columns on decimal point
\usepackage{bm}% bold math
\usepackage{subfigure}
\usepackage{color}
\usepackage[usenames,dvipsnames]{xcolor}

\usepackage{amssymb,amsmath,amsfonts,latexsym,graphicx,verbatim,dsfont}
\usepackage[matrix,frame,arrow]{xy}
\usepackage{epsf}

\usepackage[margin=0.6in]{geometry}

\usepackage[english]{babel}
\usepackage{times}
\usepackage{latexsym}
\usepackage{fancyhdr}
\usepackage{float}
\usepackage{url}
\usepackage{afterpage}
\usepackage{enumitem}
\usepackage{eso-pic, graphicx}

\definecolor{Ablue}{rgb}{0.96,0.24,0.00}

\definecolor{Abluetitle}{rgb}{0.,0.24,0.51}
\newcommand{\bluetitle}{\color{Abluetitle}}

\definecolor{orange}{rgb}{0.96,0.24,0.00}

\definecolor{darkred}{rgb}{0.55, 0.0, 0.0}

\usepackage[colorlinks=true , citecolor=blue,urlcolor=blue]{hyperref}

\input{Commands3}

\newcommand{\affA}{Department of Chemistry, University of California, Berkeley, California 94720, USA.}
\newcommand{\affF}{Materials Science Division Lawrence Berkeley National Laboratory, Berkeley, California 94720, USA.}
\newcommand{\affB}{Department of Physics, CUNY-City College of New York, New York, NY 10031, USA}
\newcommand{\affC}{CUNY-Graduate Center, New York, NY 10016, USA}
\newcommand{\affD}{Department of Chemical and Biomolecular Engineering, University of California, Berkeley, California 94720, USA.}
\newcommand{\affE}{Centre for Surface Chemistry and Catalysis, Department of Microbial and Molecular Systems (M2S), KU Leuven, Celestijnenlaan 200F P.O. Box 2461, 3001 Leuven, Belgium.}
\begin{document}
% Title & Authors
\title{\bluetitle{Wide dynamic range magnetic field cycler: Harnessing quantum control at low and high fields}}

\author{A. Ajoy}\thanks{Equally contributing authors}\affiliation{\affA}\affiliation{\affF}\email{ashokaj@berkeley.edu}
\author{X. Lv}\thanks{Equally contributing authors}\affiliation{\affA}\affiliation{\affF}
\author{E. Druga}\affiliation{\affA}
\author{K. Liu}\affiliation{\affA}\affiliation{\affF}
\author{B. Safvati}\affiliation{\affA}
\author{A. Morabe}\affiliation{\affA}
\author{M. Fenton}\affiliation{\affA}
\author{R. Nazaryan}\affiliation{\affA}\affiliation{\affF}
\author{S. Patel}\affiliation{\affA}
\author{T.F. Sjolander}\affiliation{\affA}
\author{J. A. Reimer}\affiliation{\affD}\affiliation{\affF}
\author{D. Sakellariou}\affiliation{\affE}
\author{C. A. Meriles}\affiliation{\affB}\affiliation{\affC}
\author{A. Pines}\affiliation{\affA}\affiliation{\affF}

\begin{abstract}
We describe the construction of a fast field cycling device capable of sweeping a 4-order-of-magnitude range of magnetic fields, from $\scriptsize{\sim}$1mT to 7T, in under 700ms. Central to this system is a high-speed sample shuttling mechanism between a superconducting magnet and a magnetic shield, with the capability to access arbitrary fields in between with high resolution. Our instrument serves as a versatile platform to harness the inherent dichotomy of spin dynamics on offer at low and high fields -- in particular, the low anisotropy, fast spin manipulation, and rapid entanglement growth at low field as well as the long spin lifetimes, spin specific control, and efficient inductive measurement possible at high fields. Exploiting these complementary capabilities in a single device open up applications in a host of problems in quantum control, sensing, and information storage, besides in nuclear hypepolarization, relaxometry and imaging. In particular, in this paper, we focus on the ability of the device to enable low-field hyperpolarization of  $\Cs$ nuclei in diamond via optically pumped electronic spins associated with Nitrogen Vacancy (NV) defect centers.  
  
\end{abstract}

\maketitle

\section{Introduction}
The last few decades have witnessed rapid strides in high-field superconducting magnet technology, with fields $B_0>20$T and inhomogeneities better than 1ppb routinely available, fueling several recent advances in biomolecular nuclear magnetic resonance (NMR)~\cite{Petkova202,Cady10}. In parallel, there has been a silent revolution in the development of magnetic shielding technology~\cite{Kornack07}, with specialized alloys of mu-metal providing shielding factors $>$10$^6$, and extinguishing fields to $<$0.1nT in a relatively large volume~\cite{mcdermott02,Blumich09,Ledbetter11}.

From a physical point of view, both extremes of ultra-high and ultra-low magnetic fields provide uniquely complimentary advantages. In quantum information science, for instance, high fields provide a  means to store and protect quantum information due to long spin relaxation times ($T_1$). In particular, the electronic spin associated with the nitrogen-vacancy (NV) center in diamond~\cite{Jelezko06} -- which has emerged as a promising platform for quantum information processing~\cite{Yao12,Hensen15,klimov15}, simulation~\cite{Cai13,Ajoy13b} and metrology~\cite{Maze08,Balasubramanian08} -- has a $T_1$ approaching 10ms at ~8T~\cite{Takahashi08}. In addition, high fields enable the ability to apply highly frequency selective  quantum control often with $<1$ppm resolution; as well as sensitivity gains in measurement, especially bulk inductive spin readout, where SNR scales favorably, $\propto B_0^{7/4}$~\cite{Hoult1978}. Ultra-low to zero fields (1nT-100mT), on the other hand, provide the alternative advantages of spin \I{indistinguishability} -- spins even of completely different species act identically, allowing access to heteronuclear spin singlets with long lifetimes~\cite{Emondts14} and the easy construction of Hamiltonian models in naturally occurring spin networks. More specifically, the low field regime is interaction dominated, where the spin Larmor frequencies are smaller that couplings. Indeed at zero field, with the absence of any field $B_0$ that acts to truncate the inter-spin couplings, the interaction Hamiltonians for free evolution are completely isotropic, without orientational dependence~\cite{Weitekamp83,Pines92}. This can allow the relatively easy production of nearest-neighbor Heisenberg models $\mH = \sum_j J_j \vec{S}_j\cdot\vec{S}_{j+1} + BS_{zj}$ in large spin networks in liquids~\cite{Burgarth17}. In dipolar coupled solids, this leads to fast entanglement generation since there are no disallowed transitions from energy costs set by $B_0$~\cite{Krojanski04}. Finally low-field also allows the possibility of ultra-fast quantum control, since there are no speed limits set by the rotating wave approximation.

\begin{figure*}
  \centering
	{\includegraphics[width=\textwidth]{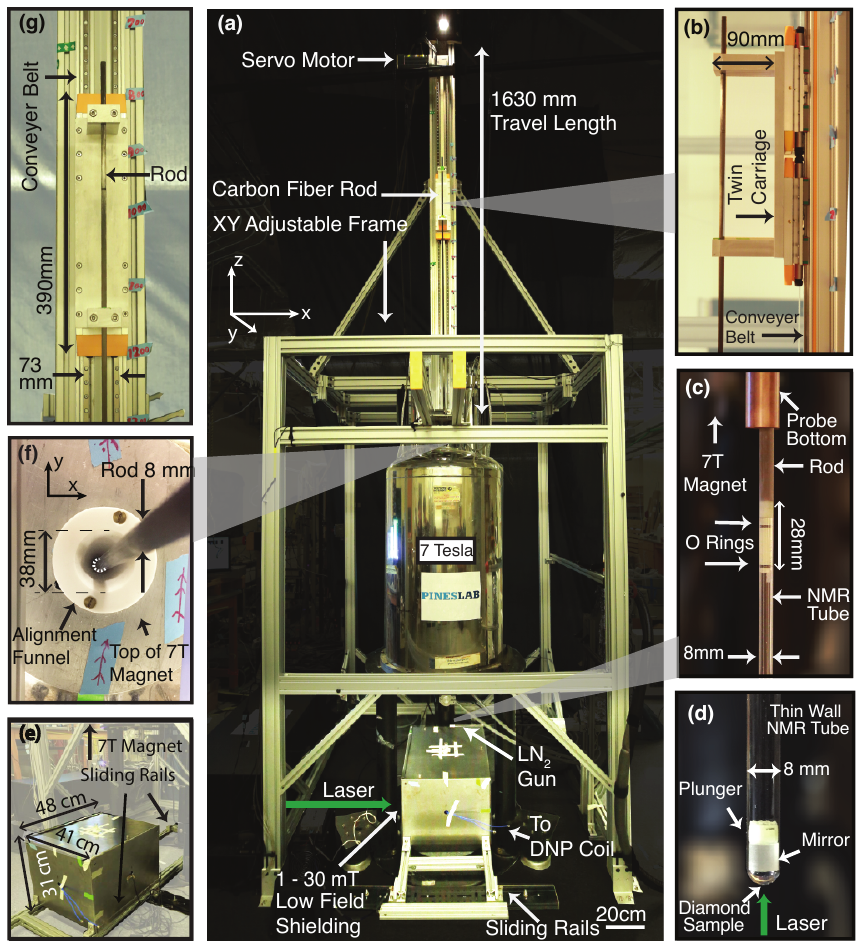}}
        \caption{\textbf{Overview of instrument.} (a) 
Mechanical shuttler constructed over a high field (7T) superconducting magnet on X and Y adjustable rails for control over alignment. A low field shield is positioned below the magnet. For hyperpolarization experiments in diamond, optimal pumping using laser and polarization transfer with microwaves to $\Cs$ in diamond particles occurs within the shielded region, after which the sample is shuttled rapidly for measurement at 7T. A liquid nitrogen (LN$_2$) gun enables rapid in-situ sample freeze. (b,g) A twin carriage actuator shuttles a carbon fiber rod along a conveyer belt. The actuator has a positional precision of 50$\mu$m and operates at a maximum speed of 2m/s and acceleration of 30m/s$^2$. (c) Panel illustrates pressure fitting using a pair of O-rings of a 8mm thin wall NMR tube containing the sample to the shuttling rod. (d) A dielectric mirror above the diamond sample increases the efficiency of optical excitation. (e) Panel details the setup within the low field shield (see \zfr{DNP-setup}). The iron shield is secured on sliding rails to prevent movement from the  magnetic force of the 7T magnet. (f) The bore of the 7T magnet is sealed with a teflon guide that allows the shuttling motion to self-align (see \zfr{guide-shuttling}).}
 \zfl{setup}	
\end{figure*}

In this work we construct a device that combines and harnesses the power of both regimes in a single platform (see \zfr{setup} and video at Ref.~\cite{shuttlervideo}) capable of sweeping magnetic fields over a four order-of-magnitude dynamic range from $\scriptsize{\sim}$1mT to 7T, and extendable to a 1nT-7T range.  The device works by physically transporting (\I{shuttling}) a sample precisely and at high speed (under 700ms) between low and high field (7T) centers placed 830mm apart, exploiting a high mechanical precision ($50\mu$m) to achieve arbitrary tunable fields in the fringing field between the two centers. The sample shuttling takes place faster than the $T_1$ times of nuclear spins in a variety of physical systems, which coupled with high resolution inductive detection at 7T make the system ideally suited to studying nuclear spins under different field environments. The system also provides the ability for spin manipulation at the low and high field centers, and rapid in-situ sample freeze to enhance $T_1$ lifetimes.

This capability paves the way for several versatile applications of the device. In this paper we shall particularly focus on applications for the quantum system consisting of coupled $\Cs$ nuclear spins and NV center electrons in diamond. Coupled to the optically addressable NV center qubit, $\Cs$ spins have garnered attention as forming viable nodes of a quantum information processor~\cite{Dutt07,Neumann10,Reiserer16} due to their long lifetimes, and the fact that they can be rapidly and directly manipulated by the NV center~\cite{Khaneja07,Borneman12,Taminiau12}. Their utility as quantum memories have been demonstrated to provide wide gains in quantum sensing, both with respect to sensing resolution~\cite{Laraoui13,Ajoy15,Rosskopf17} and sensitivity~\cite{Ajoy2016} -- especially compelling for nanoscale MRI experiments at high fields~\cite{Aslam17}. Exploiting the complimentary advantages of low and high field control that our instrument offers will enable enhanced resolution gains in ancilla assisted quantum sensing. Moreover low fields serve to strongly mitigate inherent anisotropies in the system Hamiltonians. We exploited this recently, employing our instrument,  to develop the first method for efficient room temperature dynamic nuclear polarization (DNP) of $\Cs$ nuclei via the optically pumped NV centers. The polarization transfer at low fields is orientation independent, allowing $\Cs$ hyperpolarization in \I{powdered}, randomly oriented, micro- and nano-diamonds~\cite{Ajoy17}, allowing the possibility of a ``nanodiamond polarizer'' for the optical hyperpolarization of liquids brought in contact with these high surface area diamond particles.  Sample shuttling to high fields after DNP allows an efficient detection of the $\Cs$ hyperpolarization.%This opens up the compelling possibility of a ``nanodiamond polarizer'' for the optical hyperpolarization of liquids brought in contact with these high surface area diamond particles -- the polarization mediated via spin diffusion from the $\Cs$ nuclei. 

In this paper, we focus on the instrumental (see \zfr{setup}) capabilities at low and high fields enabled by our device, that enabled the hyperpolarization experiments in Ref.~\cite{Ajoy17} . \zsr{construction} and \zsr{probe} describe the construction and design aspects of our instrument. \zsr{low-field} demonstrates our capability for low field spin manipulation, and its application for the optical hyperpolarization of $\Cs$ in diamond powder in \zsr{diamond}. \zsr{cryo} describes a cryogenic system that delivers in-situ rapid sample cooling at low field. Finally \zsr{toolbox} briefly outlines potential experiments harnessing the power of low and high fields enabled by our field cycling device.

\section{Mechanical field cycling from low field to 7T}
\zsl{construction}

The field cycler consists of a tower constructed over a high field (7T) superconducting magnet with a magnetic shield positioned below it (see \zfr{setup}). A video showing the working of the instrument is available online~\cite{shuttlervideo}. A fast conveyor belt actuator stage (Parker HMRB08) carries the sample in the fringing field of the magnet and into the shield, allowing a $\scriptsize{\sim}$1mT -7T field sweep (in principle 1nT-7T with mu-metal shields). The sample is carried by a carbon fiber shuttling rod (Rock West composites) that is fastened rigidly on a twin carriage mount on the actuator stage (\zfr{setup} a,b and g). Carbon fiber was chosen because of its exceptional strength, low mass and immunity to eddy-current forces~\cite{Redfield03}, while the twin-carriage minimizes yaw and aids in alignment, crucial for high sample filling factors.
  
\zfr{shuttler-charac} highlights the versatile control available in our system -- the ability to tune the shuttler velocity and acceleration (\zfr{shuttler-charac}a) and spatial position for start and stop of motion (\zfr{shuttler-charac}b), and motion trajectories (see \cite{shuttlervideo}).  Shuttling is possible upto a speed 2m/s and acceleration 30m/s$^{2}$ over a 1600mm travel range, with a high positional precision of $50\mu$m. This control could be exploited in systems with specific level anti-crossings (LACs); for instance in SABRE based DNP with parahydrogen~\cite{Adams09}, where we can precisely control the rate of passage through the LACs to optimize polarization transfer efficiency (\zsr{toolbox})~\cite{Theis16}. We have characterized the shuttling time from 7T-8mT to be $648\pm $2.6ms (see \zfr{shuttler-charac}a inset). This highlights the remarkably high repeatability in our instrument, which contrasts to conventional pneumatically controlled field cyclers. To our knowledge, this is also the first time sample shuttling times have been quantified with such high precision.

\begin{figure}[t]
  \centering
  {\includegraphics[width=0.49\textwidth]{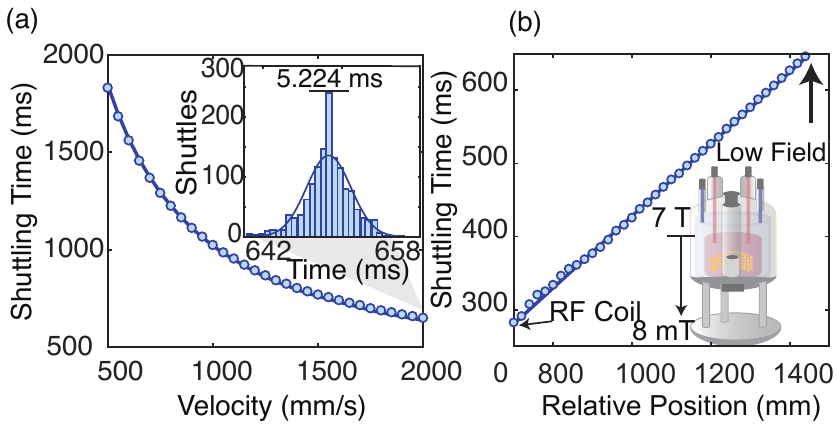}}
  \caption{\textbf{Field cycling repeatability and control.} {{(a)} Shuttling time from 8mT where the DNP excitation occurs to 7T as a function of shuttling velocity, where times are measured from the arrival of a trigger pulse from the actuator (see \zfr{trigger-sequence}).  The inset demonstrates the high repeatability of the system, with travel time 648$\pm 2.6$ms over 1400 trials shuttler operation. {(b)} Measurements of shuttling time as a function of distance from the NMR coil at 7T, revealing the high mechanical precision of the instrument.} }
\zfl{shuttler-charac}	
\end{figure}

\subsection{Shuttling alignment and vibrational stability} 
\zsl{alignment}
Our system incorporates special features to maximize the sample filling factor for highly efficient inductive detection, and for low field quantum control through radiofrequency or microwave excitation (\zsr{low-field}). For minimum possible clearance to excitation coils at both fields, and low vibration associated with motion jerk, it is essential that the shuttling rod be aligned parallel to the magnet over the entire distance of travel. We align to better than 1mdeg through a series of design implementations. Firstly, the entire shuttling tower (80/20 1530-S) containing the actuator, motor and twin carriage is on an XY tunable platform (\zfr{setup}a, and video at Ref.~\cite{shuttlervideo}). This centers the shuttling rod to the magnet bore with a precision better than 0.25 mm over the 1600 mm travel. Secondly, and more critically, two alignment funnel-shaped guiding stages made of soft teflon are employed at the magnet bore (\zfr{setup}f), and NMR probe (\zfr{probe}b).  The stages vertically align the structure, and provide additional points of support to greatly reduce vibration (see video at Ref.~\cite{shuttlervideo}). The carbon fiber shuttling rod (8mm diameter, 1.7m length), while soft enough to be guided by the teflon stages, is inherently less prone to vibration due to its low moment of inertia and high strength (430 GPa tensile modulus). 

\zfr{guide-shuttling} visualizes the guiding process, taken with a camera located in the NMR probe focusing on the teflon guide above it (\zfr{probe}b). The rod starts out slightly misaligned but is \I{dynamically} guided to be perfectly aligned, the carbon fiber malleable enough to be able to guide into place with no jerk. The funnel guide has a 45 degree taper with an opening of 8.077 mm, ensuring a tight fit with the 8mm shuttling rod.  This ensures shuttling with minimum clearance to the NMR coil and consequently high filling factors (see \zfr{shuttling-movie}). 

\begin{figure}[t]
  \centering
  {\includegraphics[width=0.49\textwidth]{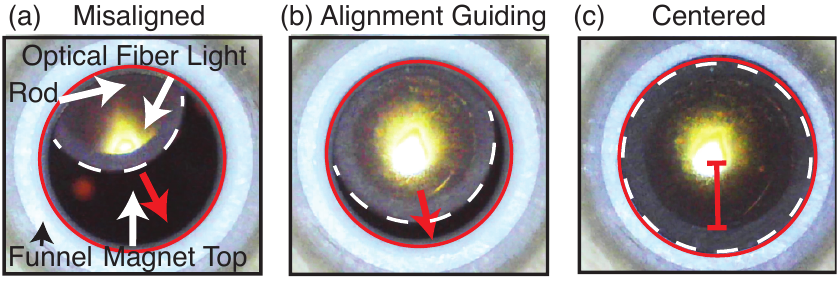}}
  \caption{\textbf{Guided self-aligning high speed shuttling.} Panels illustrate the guiding action of the teflon funnel right above the RF coil that allows for self-aligned high speed shuttling within a misalignment of less than 1 mdg. For illustration purposes we employ an optical fiber light inside the carbon figure shuttling rod, and a camera mounted inside the NMR probe (see also \zfr{shuttling-movie}). {(a)} The rod approaches the funnel misaligned, but is dynamically guided to be centered (panels {(b-c)}).}
\zfl{guide-shuttling}	
\end{figure}

\subsection{Sample attachment to shuttling rod}

The sample is pressure-fit to the carbon fiber rod for rapid attachment and detachment (see video at Ref.~\cite{shuttlervideo}). The lower end of the rod contains a ceramic connection for attaching the NMR tube carrying the sample (\zfr{setup}c). It consists of a pair of soft, high temperature modulus, silicone O-rings (McMaster 1/16 Fractional Width, 0.254" OD). Remarkably, this arrangement proves resilient for fast shuttling with just the simplicity of a hand-tight pressure fit. The diameter of the NMR tube (Wilmad 8mm OD, 1mm thickness) was chosen to match that of the shuttling rod for a seamless joint through the alignment guides.

\zfr{setup}d details the tube containing a sample of powdered diamond employed in optical hyperpolarization experiments. A plunger carrying a dielectric mirror is used to isolate the sample to a compact volume. The plunger is fitted with a threaded screw hole for easy fastening access, positioning, and removal. The mirror (Thorlabs BB1-E02 Broadband Dielectric Mirror, 400 - 750 nm) is machine ground to the inner diameter of the NMR tube, and provides a double pass for the incoming laser radiation for efficient polarization of NV center electrons.

\begin{figure}[t]
  \centering
  {\includegraphics[width=0.49\textwidth]{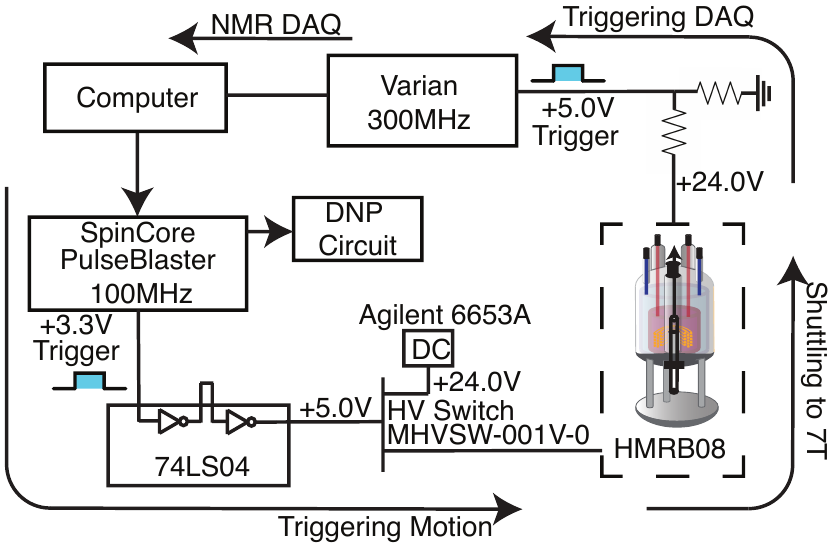}}
  \caption{\textbf{Trigger sequence for field cycling and detection.} The actuator motion is triggered by a 24V pulse indirectly generated from the PulseBlaster controller. For instance, in experiments for optical polarization transfer to $\Cs$ spins in diamond (\zfr{toolbox}a), the shuttling motion occurs from the low-field region where DNP is excited ($\sim$ 8mT) and finally to high-field 7T, after which the actuator returns a pulse that triggers the NMR measurement.
}
\zfl{trigger-sequence}
\end{figure}

\subsection{Synchronization and Triggering}
\zsl{sync}

Shuttling and inductive detection are synchronized (see \zfr{trigger-sequence}) using a high-speed pulse generator (SpinCore PulseBlaster USB 100 MHz). The servomotor is triggered to start motion for shuttling with a 24V 10ms pulse. This is generated by upconverting the 3.3V pulse trigger from the pulse generator to 5V by a TI SN74LS04N CMOS inverter, which then serves as the gate to a high voltage MOSFET switch (Williamette MHVSW-001V-036V). The MOSFET switch relays the 24V pulse to the servomotor that drives the belt-driven actuator stage to the desired position. Finally, the actuator returns a 24V pulse indicating the completion of motion, which is passed through a voltage divider to trigger the NMR console (Agilent DD2) to initiate measurement (see \zfr{trigger-sequence}). This is also used to precisely quantify the shuttling times and jitter (\zfr{shuttler-charac}).

\subsection{Comparison with other field cycling platforms}
Let us now briefly compare our field cycling platform to previous reports. Unlike conventional field cycling platforms, that are geared towards relaxometry experiments in liquids and proteins, our device seeks to harness the versatile dichotomy in spin dynamics regimes between low and high fields.  Our instrument allows a large dynamic range magnetic field sweep from 1mT-7T, which can be further extended to a ten order-of-magnitude range 1nT-7T through enhanced low field shielding, following a zero-field setup similar to the one described in Ref.~\cite{Tayler17}. Note that in contrast, fast field cycling~\cite{Lips01} which uses specialized power supplies and switched-coils ~\cite{Sousa04,Ferrante05} to rapidly switch between magnetic fields, cannot easily access fields $>$2T. For experiments described in \zfr{toolbox}, the ability to manipulate and store spins at higher field enables applications not accessible by pulsed field cycling.

Focusing now on field cycling devices employing sample shuttling, they come in two flavors: pneumatic and mechanical shuttling. Pneumatic devices generally provide higher shuttling speeds even though it is challenging to measure the speeds precisely; however compressed air causes sample vibration upon motion start and stop, which adds an additional 0.1-1s time for stabilization. Our mechanical instrument operates with high positional precision (50$\mu$m), shuttles with a maximum velocity of 2m/s with high repeatability (\zfr{shuttler-charac}), and has a large thrust force capacity (295N). While the lower speed limits it from certain tasks (eg. protein relaxometry), the precision of control over position and velocity makes it ideal for a plethora of other applications (\zsr{toolbox}), for instance for field sweeps through energy level anti-crossings over a wide dynamic range. 
  
While inspired by the pioneering mechanical field cycling platform of Redfield~\cite{Redfield12},  our device extends his original innovative field cycler in several directions. First the use of a more precise actuator stage allows the ability to access fields at very high resolution, and with high repeatability (\zfr{shuttler-charac}). The fact that the low field center is far separated from the highfield center, and primarily situated at the bottom of the magnet, allows  one to accessorize the field cycler with optical and microwave irradiation and cryocooling (see \zsr{cryo}) components for spin control at low fields. This makes the field cycler as suitable for optical spin hyperpolarizaton experiments (\zsr{diamond}) involving the nitrogen vacancy center in diamond. In contrast to other mechanical shuttling designs~\cite{Chou12}, the shuttling assembly occupies only 8mm in the magnet bore, allowing interfacing with magnetic inserts~\cite{chou15} which would allow the creation of homogeneous intermediate field regions for quantum control~\cite{Cousin16}.

\begin{figure}[t]
  \centering
  {\includegraphics[width=0.38\textwidth]{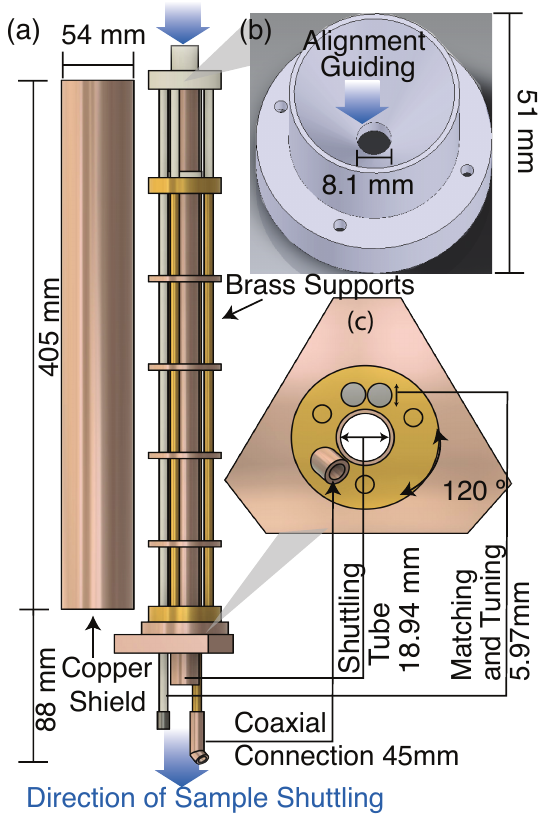}}
  \caption{\textbf{NMR probe for high speed shuttling} which enables high speed shuttling from high to low fields. (a) The probe has a hollow Cu-shielded cylindrical opening through which the carbon fiber rod carrying the sample is shuttled. (b) The top of the probe consists of a teflon funnel guide that helps in aligned shuttling (\zfr{guide-shuttling}). Either split or saddle coils (\zfr{DNP-setup}) can be used for $\Cs$ NMR detection at 75 MHz. }
\zfl{probe}
\end{figure}

\begin{figure}[t]
\centering
  {\includegraphics[width=0.49\textwidth]{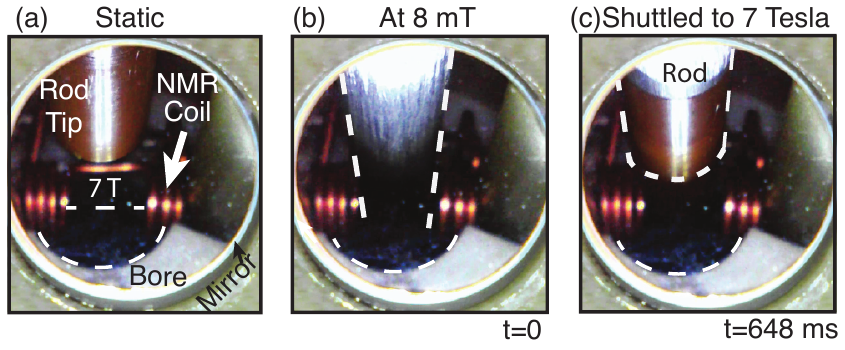}}
  \caption{\textbf{Snapshots of high-speed sample shuttling}{ obtained with a camera mounted inside the NMR probe in the 7T superconducting magnet bore. Note that the rod tip  here was redesigned with a Vespel holder for single crystal samples. The camera is focused on a mirror mounted on the bottom face of the teflon funnel that guides the shuttling (\zfr{probe}). {(a)} Sample approaching split NMR detection coil. {(b)}  Shuttling to the low-field center, for instance for DNP excitation. The shuttling time is 648$\pm 2.6$ms (\zfr{shuttler-charac}). {(c)} Sample approaching the center of the NMR coil with better than 100$\mu$m precision.  Clearance to the sides of the coil is under 0.5 mm. }}
 \zfl{shuttling-movie}	
\end{figure}

\begin{figure}[t]
  \centering
  {\includegraphics[width=0.49\textwidth]{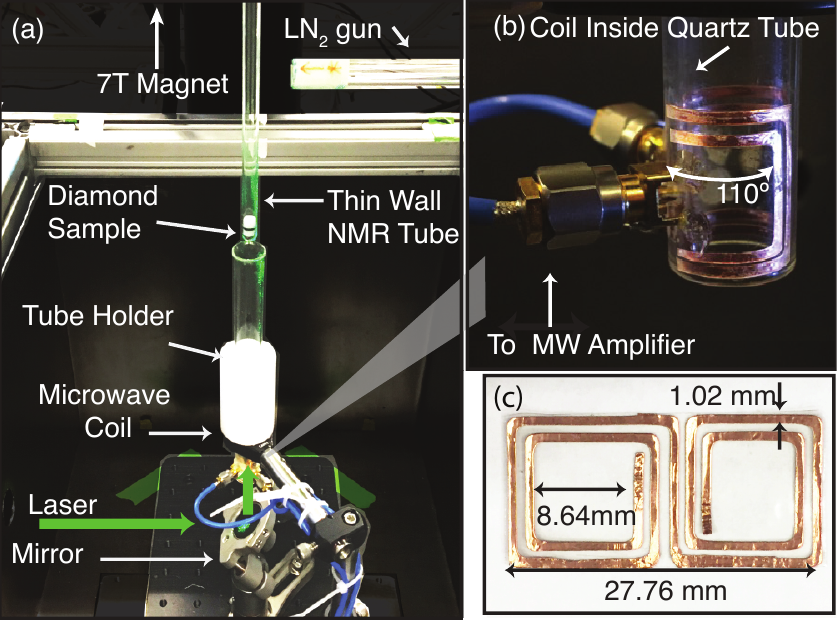}}
  \caption{\textbf{Low field optical hyperpolarization setup.} {Figure illustrates the setup for polarization transfer at low field ($\sim$ 1-30 mT), performed inside a magnetic shield under the 7T NMR magnet (see video at Ref.~\cite{shuttlervideo}). {(a)} The carbon fiber rod attached to the NMR tube carrying the diamond sample is shuttled into a DNP excitation coil. Laser irradiation (5W over a 8 mm beam diameter) is applied from the bottom of the tube with a 45 degree mirror. {(b)} Overview of the DNP coil wrapped on the inside of a quartz tube and showing connections to the 100W microwave (MW) amplifier. {(c)} Panel shows the coil originally printed on copper foil for a 9mm diameter tube. (see ~\cite{coilvideo2}).}}
\zfl{DNP-setup}	
\end{figure}

\section{NMR probe compatible with shuttling}
\zsl{probe}

Modifications were made to conventional NMR probe design to accommodate fast sample shuttling (\zfr{probe}) \I{through} it. \zfr{shuttling-movie} visualizes the actual shuttling process at the probe. The probe is designed hollow for shuttling to low fields below the magnet and is constructed out of 12.7 mm thick brass plates for enhanced shielding and structural rigidity. The top plate holds the tuning and matching capacitors (Voltronics AT4HV and AP14) along with the quartz tube (ID 9 mm $\zt\:$ OD 11 mm Technical Glass Products) outside of which is fabricated a saddle shaped NMR coil. The probe can accommodate either split solenoid coils (\zfr{shuttling-movie}) or current saddle shaped ones (similar to \zfr{DNP-setup}c).  The teflon funnel at the top (\zfr{probe}b) allows for the rod to self align (\zsr{alignment}). Assuming the sample fills the full height of the coil, the filling factor is $\app$0.4, which is comparable to commercial probes, allowing high sensitivity inductive detection. 

\subsection{Saddle coils for maximizing filling factor}
\zsl{coil}

We have developed a rapid technique of fabricating saddle coils for NMR detection that are compatible with shuttling, yet optimized to the sample of interest to provide large filling factors. A descriptive video showing the fabrication method is available online~\cite{coilvideo2}.  The coils are wrapped on a Quartz tube matched to the NMR tube being shuttled such that the sample tightly fills the entire coil volume. The coils are fabricated by cutting them out of flexible adhesive copper foil (Venture Tape, thickness 31.75 $\mu$m) (see \zfr{DNP-setup}c) using an inexpensive vinyl cutter (Silhouette Cameo2) in 17 seconds and 1.5mm cutting depth (see video at Ref.~\cite{coilvideo2}). The RF coil in the probe has 10.56mm 110$^{\circ}$ windows~\cite{Hoult1978} and track width of 1.25 mm with 0.75 mm spacing. After printing, excess copper around the coil was carefully removed and contact paper (Circuit StrongGrip Transfer Tape) was applied on top to maintain the coil shape (see ~\cite{coilvideo2}). When inserted into the quartz tube, the inside of the tube was coated with water to prevent the coil from adhering while positioning to a slit for the leads to be pulled through. A heat gun was used to release the contact paper and adhere the coil to the wall of the tube. The fabricated coils for $\Cs$ NMR had an inductance of  0.28$\mu$H and Q factor of 150 at 75.03 MHz.

\begin{figure}[t]
  \centering
  {\includegraphics[width=0.35\textwidth]{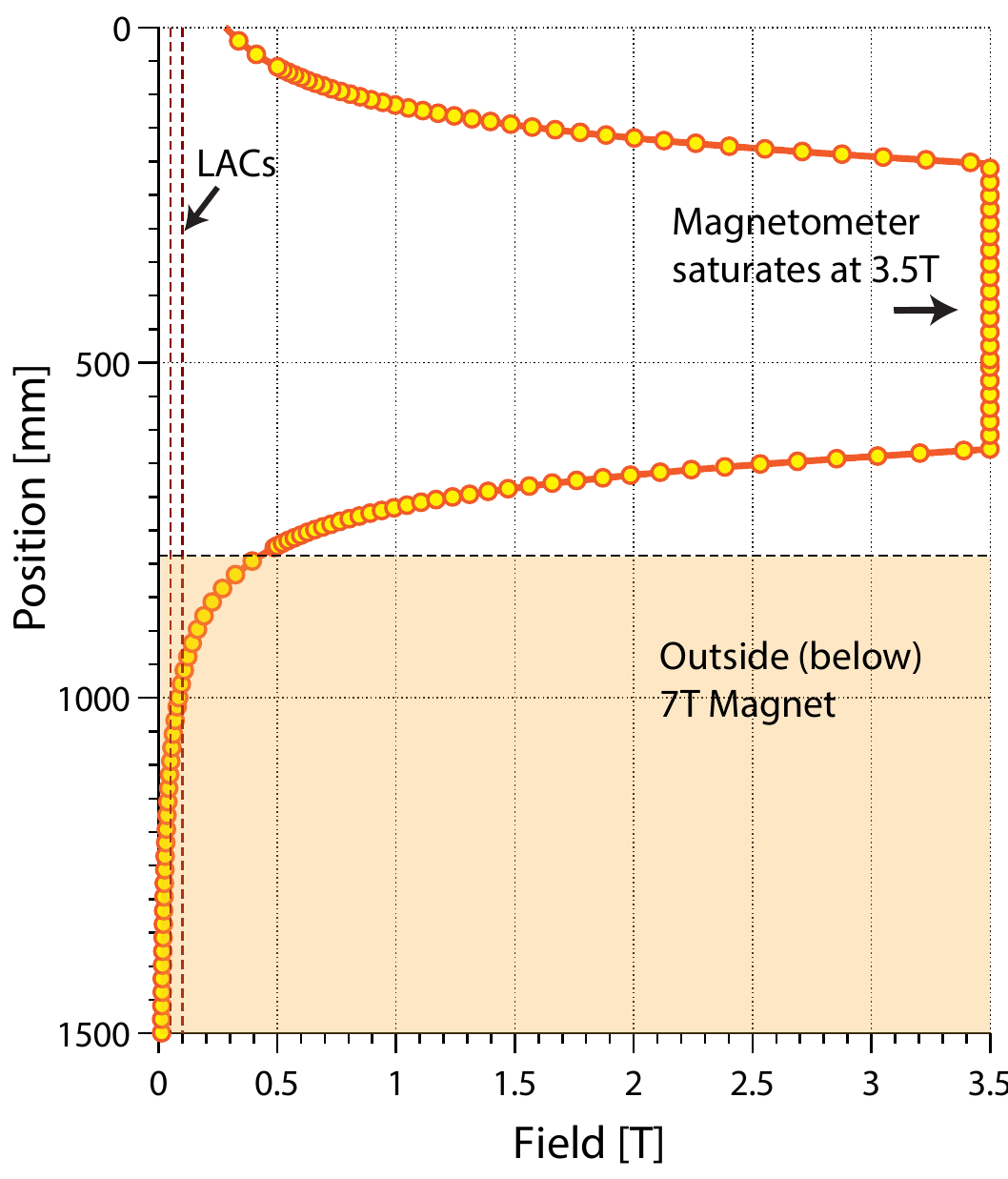}}
  \caption{\textbf{Full field map} characterizing the fields natively accessible in our field cycler. Note that we measure the longitudinal $B\hat{\T{z}}$ component of the field, and for these measurements we have removed the low-field shield to just illustrate the native fringing field of the magnet (fields $>$3.5T are not delineated due to saturation of the magnetometer). The addition of the shielding allows the modification of the low field characteristics with negligible effect at high field. The instrument is hence able to access a wide field range with high field resolution, set ultimately by the 50$\mu$m sample positional precision (see \zfr{LAC})}
\zfl{field-map}	
\end{figure}

\begin{figure}[t]
  \centering
  {\includegraphics[width=0.49\textwidth]{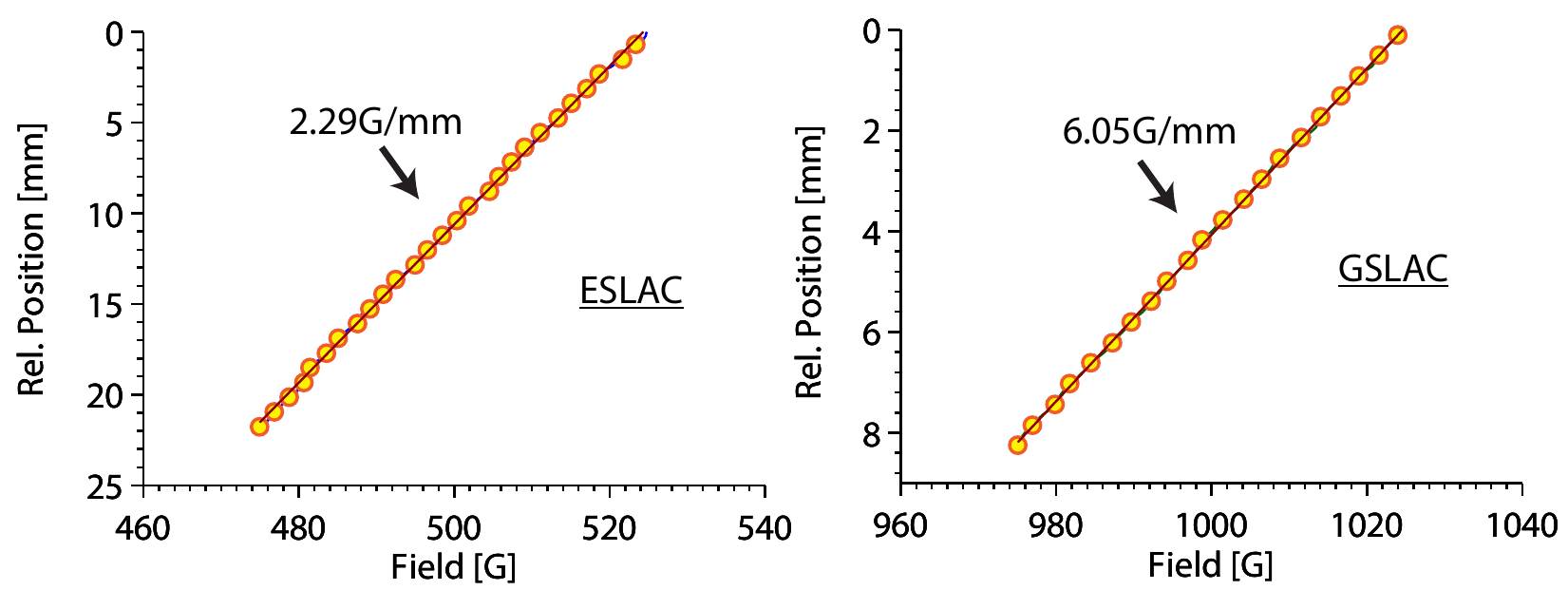}}
  \caption{\textbf{Accessing level-anticrossings in diamond.} Figure illustrates the spatial longitudinal field dependence at shuttling positions corresponding to the diamond NV center (a) ESLAC $\app$ 510G and (b) GSLAC $\app$ 1020G. The gradients are linear to a good approximation. (a) At the ESLAC, the 50$\mu$m positional precision of our instrument allows a field resolution of 0.114G and maximum sweep rate of 0.458T/s considering a speed of 2m/s. (b) Our shuttler can sweep through the GSLAC at 1.21T/s with a resolution of 0.303G.}
\zfl{LAC}	
\end{figure}

\begin{figure}[t]
  \centering
  {\includegraphics[width=0.49\textwidth]{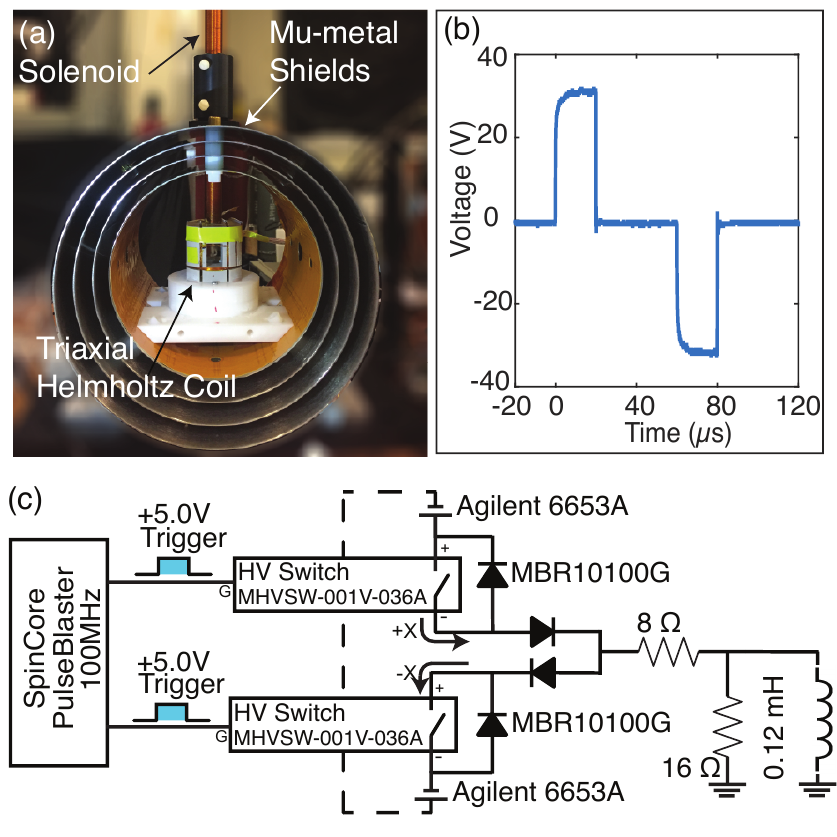}}
  \caption{\T{Zero-field center and quantum control.} Enhanced shielding installed at the low-field center allows access to the zero field regime~\cite{Tayler17}.  (a) Cross section of zero-field ($\app 1$nT) center consisting of concentric mu-metal shields designed to fit the low-field iron shield (\zfr{setup}e). A solenoid leading into the shield enables implementation of a Hamiltonian \I{quench} to zero-field. Quantum control at zero-field is implemented via fast DC magnetic pulses (shown in (b)) generated within a triaxial Helmholtz coil is mounted on a 3D printed coil holder. (c) Switching circuit that generates fast DC pulses for spin manipulation at zero-field, shown here for a single channel (X,-X). MOSFET switches triggered and synchronized by Pulse Blaster switch current into the coils from a high-current power supply (Agilent 6653A). We routinely achieve ~1$\mu$s pulse rise times with $\app$ 4A of current (see (b))}
\zfl{ZF-setup}	
\end{figure}

\section{Low-field center}
\zsl{low-field}
\subsection{Construction}
Our field cycling device provides a wide dynamic field range by employing a magnetically shielded low-field location that the sample can be shuttled into (\zfr{setup}e and video at Ref.~\cite{shuttlervideo}). Since the shields sit in the $\approx$300G fringing field of the 7T magnet, they are positioned on sliding rails to secure them against upward magnetic forces. The shield is constructed out of concentric layers of stress annealed iron (NETIC S3-6 alloy 0.062'' thick, Magnetic Shield Corp.) and mu-metal (TwinLeaf MS-1), with the iron on the outside due to its high saturation, and mu-metal on the inside due to its high permeability (over $10^6$ with 4 layers). In combination, with 3 layers of iron and 4 layers of mu-metal, one can achieve a lower field center of approximately 1nT.  In practice, the low-field shielding can be customized to suit the target field desired in experiments, in particular -- low fields (1-30mT) for optical DNP experiments in diamond~\cite{Ajoy17}, ultra-low fields ($<1\mu$T) for relaxometry, and even ‘zero’ fields ($<1$nT) for applications exploiting isotropic Hamiltonians models~\cite{Burgarth17, Llor91,Appelt10,Ledbetter11} and decoherence free subspaces ~\cite{Emondts14,Pileio10} readily available at these fields.

\subsection{Field map}
We have obtained a $B_z$ field map over the full travel range of the sample from low to high field (\zfr{field-map}). Measurements were performed with a sensitive longitudinal field Gaussmeter (Lakeshore HMMA-2504-VR-10) that is inserted into the hollow shuttling rod so as to be aligned centered with the bore.  The fact that the superconducting magnet is unshielded leads to rather weak gradients in the fringing field, allowing one to access fields with $<$1G resolution over a wide range given the 50$\mu$m precision over sample position. This is highlighted in \zfr{LAC} for the excited (ESLAC $\app$510G) and ground (GSLAC $\app$1020G) state level crossings of NV centers in diamond. The weak gradients manifest as approximately linear spatial field dependencies. Indeed our instrument allows us to access the ESLAC region with a $\app$114mG resolution, and sweep through it at 0.458T/s. This would allow applications for the optical hyperpolarization of $\Cs$ nuclei and P1 centers in the vicinity of the ESLAC (\zfr{toolbox}f)~\cite{Fischer13,wunderlich17,Pagliero17}.

\subsection{Zero-field quantum control}
While our focus in this paper is the use of the device in the low-field regime, our instrument also potentially allows one to achieve a zero-field~\cite{Tayler17} ($<$1nT) center by interfacing concentric mu-metal shields into the iron low-field volume (\zfr{ZF-setup}). Quantum control at zero-field can be enabled using fast DC magnetic field pulses.  We have developed a pulsing circuit consisting of high power MOSFETs that allow the rapid switching of current from a 600W power supply generating fast DC pulses for spin manipulation (see \zfr{ZF-setup}c). The free evolution of Hamiltonian in the absence of a static magnetic field without laser and microwave irradiation is isotropic. This opens the possibility to exploit isotropic Hamiltonians and fast entanglement growth that naturally occurs at zero-field for several quantum simulation problems~\cite{Burgarth17}, while retaining the high detection sensitivity at high field. A detailed description of experiments in this regime will be presented elsewhere~\cite{Ajoy17u}.

\section{Applications to $\Cs$ optical hyperpolarization in diamond}
\zsl{diamond}
We recently employed the unique field cycling ability of our instrument, along with the ability of laser and microwave control of the NV electronic spins at low field to develop a novel method for dynamic nuclear polarization (DNP) of $\Cs$ spins in powdered diamond ~\cite{Ajoy17}. The aim of this paper is to highlight how the complimentary advantages of low and high fields enabled by our instrument enabled this advance. Low fields (typically employed $B _{ \text{pol}}\sim$1-30mT) mitigate of the strong orientation dependence of the NV centers, while high fields allow efficient detection of the generated hyperpolarization. Importantly, low fields allow one to invert the conventional hierarchy used in DNP experiments, entering the regime where the nuclear Larmor frequency is smaller than the electronic hyperfine interaction. This engenders hyperpolarization generation through microwave frequency sweeps of the electronic spectrum.

 To be more specific, \zfr{DNP-setup} describes the DNP setup in the low-field center (see video at Ref.~\cite{shuttlervideo}). The sample is irradiated simultaneously with laser and microwaves sweeping across the inhomogeneously broadened NV center powder pattern to perform the polarization transfer (\zfr{dnp}a). The microwave excitation is produced via coils (outlined in \zsr{coil}) connected to a high power (20-100W) microwave amplifier. 
%A detailed description of the mechanisms for polarization transfer are presented elsewhere~\cite{Ajoy17}.
The DNP mechanism (see \zfr{dnp}c) can be interpreted as partly-adiabatic traversals of Landau-Zener crossing in the rotating frame, with a model of NV electron coupled to a single $\Cs$ nucleus~\cite{Ajoy17}.  In the low field hyperpolarization regime, the $\Cs$ nuclear Larmor frequency $\omega_L =\gamma B_{\text{pol}}  \lesssim \left| A \right|$. In the NV sublevel $m_s=\pm 1$, the eigenstates are set by the hyperfine coupling, denoted as $\beta_{\uparrow}$ and $\beta_{\downarrow}$. On the other hand, nuclear eigenstates become dominant in the $m_s=0$, manifold, and are denoted as $\alpha_{\uparrow}$ and $\alpha_{\downarrow}$. When taking second order hyperfine coupling into consideration, Larmor frequency $\tilde { \omega } _ { L } \approx \omega _ { L } + \frac { \gamma _ { e } B _ { \text { pol } } A \sin \vartheta } { \Delta - \gamma _ { e } B _ { \text { pol } } \cos \vartheta }$, where $\vartheta$ is the orientation of N-to-V axis respect  to external magnetic field, and $\Delta$=2.87GHz is the NV zero-field splitting. Sweeping microwaves across this set of transitions, leads to their sequential excite a set of transitions that drive the coherence hyperpolarization transfer~\cite{Ajoy17, Ajoy18}.  We demonstrated that the hyperpolarization transfer is an orientation independent hyperpolarization technique, and is enabled by the relatively small inhomogeneous broadening ($\app $400MHz) of the NV electronic linewidth (powder pattern) at low fields (1-30mT). As a typical example, for 200$\mu$m diamond particles (Element6) containing about 1ppm of NV centers, the $\Cs$ polarization at a field of 13mT was enhanced to the level corresponding to thermal magnetization at a field higher than 1900T. Field cycling allows one to harness the high sensitivity of inductive measurement at 7T (\zfr{dnp}b), enabling the unambiguous inductive readout of the bulk $\Cs$ nuclear magnetization. Indeed hyperpolarization enables a measurement time gain of about five orders of magnitude at 7T, allowing the detection of a single 200$\mu$m particle with unit SNR in one shot~\cite{Ajoy17}.

The high-precision and programmable field cycling ability of our setup also proves beneficial in relaxometry, in determining the factors affecting the spin lifetimes times of the $\Cs$ nuclei in diamond.  \zfr{T1B} for instance shows a relaxation-field map $T_1({B})$ for 200$\mu$m powdered diamond microparticles at various fields ${B}$. For these measurements, we first begin by hyperpolarizing the $\Cs$ nuclei \I{opposite} to the direction set by ${B}_0=$7T. The sample is then rapidly shuttled to the target field $B_\R{relax}$, allowed to wait for time $T_\R{relax}$, and then rapidly shuttled again to 7T for detection. Sweeping $t$ and fitting the decay to a monoexponential provides the lifetime $T_1({B})$. Substantial time savings,  typically 4-5 orders of magnitude, are obtained via the hyperpolarization process.  \zfr{T1B}a shows typical relaxation data at explemary low and high fields. We measure the $T_1$ times to be 395.7s at ${B}_0=7$T and 10.19s at ${B}_{\R{pol}}=$8mT. %(see \zfr{T1B}a) for 200$\mu$m diamond particles (Element6) with an NV center concentration of $\sim$1ppm and nitrogen concentration $\app$200ppm. 
Since the DNP polarization is inhomogeneous, i.e. spins closer to the NV centers are more strongly polarized, spin diffusion between them also causes a signal decay at low fields, leading to a super-exponential fall in signal. This feature slightly underestimates the $T_1$ at ${B}_{\R{pol}}$. The field dependent relaxation data in \zfr{T1B}b shows a steep dependence with field, the spin lifetimes falling rapidly at low fields,  but growing to long room temperature polarization lifetimes ($\sim$5min) beyond 1T. This feature arises from an increasing overlap between the $\Cs$ Larmor and electronic dipolar reservoirs associated for instance with P1 centers at low fields, and will be addressed in detail in a future manuscript.

\begin{figure}[t]
  \centering
  {\includegraphics[width=0.49\textwidth]{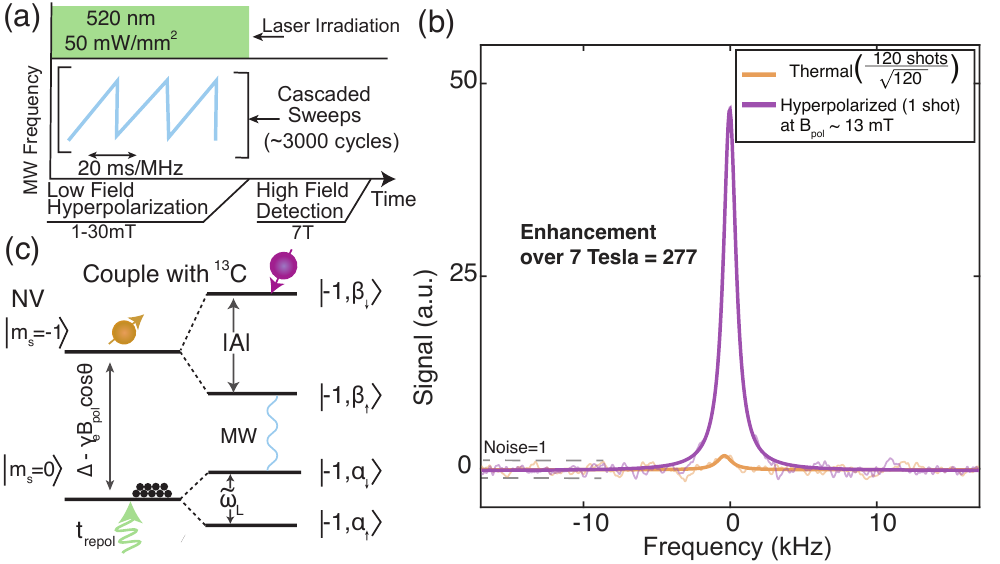}}
  \caption{\textbf{$\Cs$ optical hyperpolarization in diamond.} (a) Hyperpolarization protocol via laser irradiation that polarizes NV center to $m_s=0$ sublevel (close to $100\%$) and frequency/field sweeping over NV center ESR spectrum which transfers the polarization to $\Cs$ nuclei~\cite{Ajoy17}. (b)  $200\mu m$ diamond powder hyperpolarized at 13mT displays 277 times signal gain over Boltzmann thermal polarization. Orange line shows the $\Cs$ NMR signal due to Boltzmann polarization at 7T, averaged 120 times over 7 hours, while the purple line is a single shot DNP signal obtained with 40s of optical pumping. The signals are displayed with their noise unit-normalized for comparison. (c) Model of NV center coupled to a single $\Cs$ nucleus with hyperfine interaction $A$. The N-V axis oriented at $\theta$ respect to external magnetic field $B_{pol}$. At low polarizing fields, we access the regime where the $\Cs$ nuclear Larmor frequency $\omega_L =\gamma B_{pol}  \lesssim \left| A \right|$~\cite{Ajoy17}. } 
\zfl{dnp}	
\end{figure}

\begin{figure}[t]
  \centering
  {\includegraphics[width=0.5\textwidth]{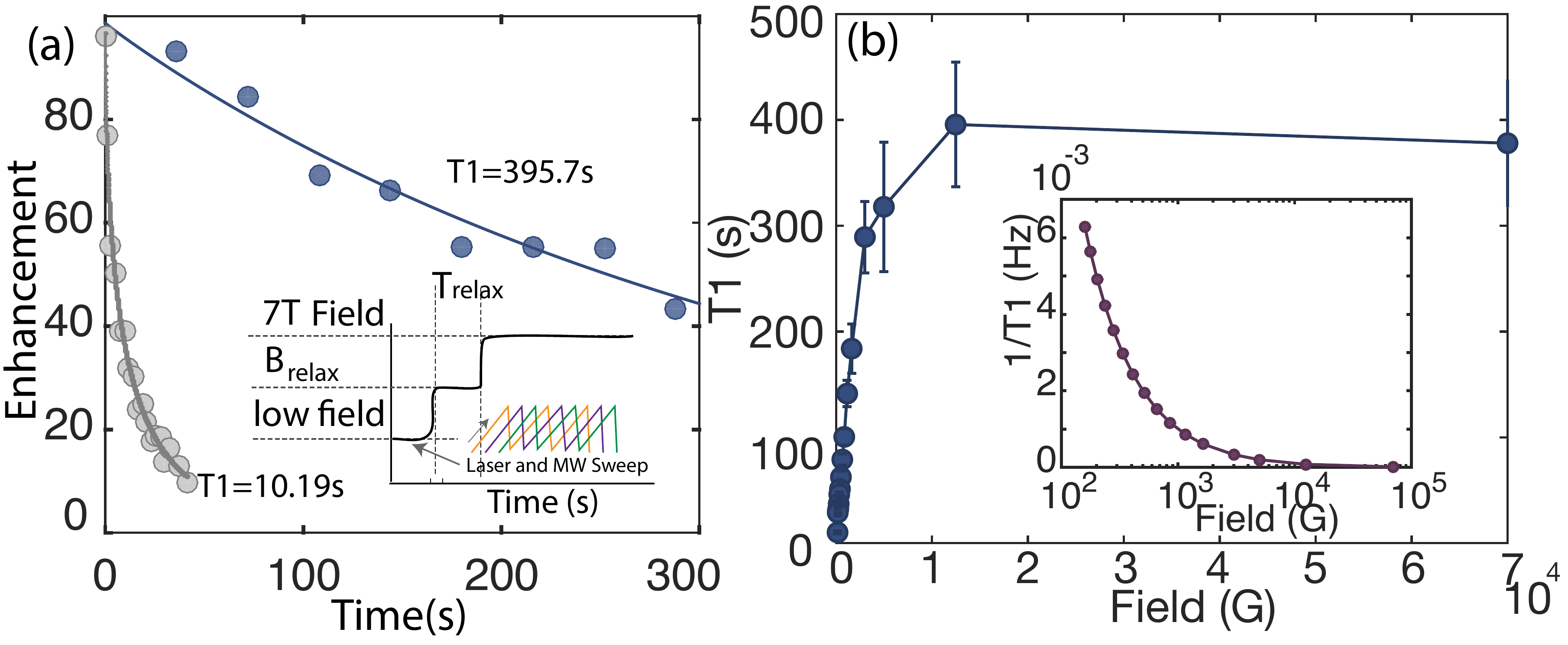}}
  \caption{\textbf{Mapping $\Cs$ spin relaxation times.} (a) $T_1$ lifetimes of $\Cs$ nuclear spins in 200$\mu$m diamond particles measured experimentally with our instrument at two explemary fields, at low field ${B}_{\R{pol}}=$8mT where DNP is excited and high field ${B}_0=$7T where the polarization is detected. The low-field $T_1$ shows a super-exponential decay, which we hypothesize arises due to spin diffusion. \I{Inset:} Measurement protocol. After optically hyperpolarized at low field, the sample was rapidly delivered to a designated $B_\R{relax}$ to relax for $T_\R{relax}$, followed by a rapid shuttling to high field and subsequent detection. (b) Map of  field dependence of relaxation time, $T_1(B)$ was obtained by extracting the monoexponential decay constant from the full decay curve at every field value. We observe a sharp increase in relaxation time $T_1$ past a field of $\app$0.5T. This dependence arises from an interaction of the $\Cs$ nuclei with the dominant dipolar coupled bath of P1 centers. \I{Inset:} Relaxation rate $1/T_1$  on log scale.}
\zfl{T1B}	
\end{figure}

\begin{figure}[t]
  \centering
  {\includegraphics[width=0.49\textwidth]{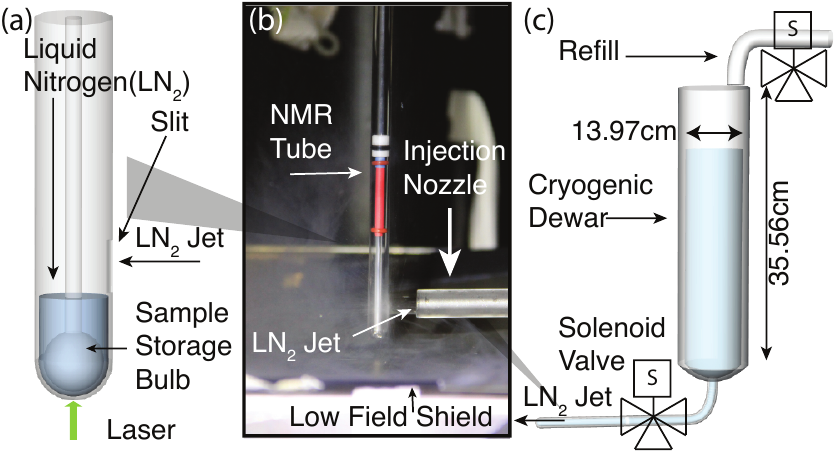}}
  \caption{\T{Cryogenic system for rapid sample freeze.} (a) An NMR tube is filled with liquid nitrogen through a fine slit to immerse the glass bulb containing the sample in a cryogenic bath. The sample is frozen uniformly within 3-4s before sample is laser irradiated from the bottom of the tube. (b) Panel demonstrates liquid nitrogen injection. The cryogenic jet flows from the quartz nozzle for 1s. Under laser irradiation, the liquid remains in the tube for 30s. (c) Panel depicts the dewar that creates constant pressure for liquid flow. The dewar is (re)filled from the top and stores 5.45L of cryogenic liquid for 1 hour. Solenoid valves are triggered by the Pulse Blaster (see \zfr{trigger-sequence}) to control refilling and ejection of liquid nitrogen.}
\zfl{LN2}	
\end{figure}

\section{In-situ fast sample freeze}
\zsl{cryo}
We have interfaced our field cycling instrument with a homebuilt liquid nitrogen (LN$_2$) cryogenic system for fast in-situ freezing of the sample at the low-field center (\zfr{setup}a). The primary aim is to increase the sample $T_1$ lifetimes. For instance, $\Cs$ spins in pyruvate, an important molecule in the metabolic cycle and cancer detection, can exceed 55s at 10mT~\cite{Chattergoon13}, but when frozen the resulting $T_1$ can be nearly an hour~\cite{van11}. A striking example of where such sample cooling would be useful is when employed for a ``\I{nanodiamond hyperpolarizer}'' -- where polarization can be transferred from optically polarized $\Cs$ nuclei in high surface area diamond particles to external $\Cs$ in the frozen solution~\cite{Ajoy17} -- the long $T_1$ times enabling a larger buildup of hyperpolarization.

\zfr{LN2} describes the setup. The design incorporates a means to rapidly cool the sample in a manner that can be synchronized with the device shuttling (\zfr{trigger-sequence}). A consistent jet of cryogenic liquid is rapidly ejected from a nozzle (see \zfr{DNP-setup}) on-demand which cools the sample to 77K (liquid nitrogen temperature) within 3-4 seconds. Flowing N$_2$ gas is applied to evacuate water vapor to prevent condensation. A dewar with two openings (\zfr{LN2}c) stores cryogenic liquid and creates the pressure that drives the LN$_2$ flow. The top opening is for (re-)filling by an external liquid nitrogen source, and the bottom for injecting liquid into a quartz nitrogen nozzle to generate a jet (\zfr{LN2}b). Both inlet and outlet are regulated by two solenoid valves individually (Asco 1/4” Cryogenic Solenoid Valve, 7/32” orifice diameter, 24VDC control voltage), which can be triggered indirectly by the Pulse Blaster and respond within 1ms to ensure synchronization (see \zsr{sync}). The system is designed so that laser irradiation and cryogenic cooling can occur simultaneously, and the LN$_2$ level in the dewar is constantly replenished to maintain a constant jet pressure.

The exact manner in which the sample cooling occurs is detailed in \zfr{LN2}a. A fine slit in the sample tube allows an inlet to the cryogenic liquid, ensuring that it forms a cold bath around the sample. The sample itself is held in a glass bulb within the tube. The liquid remains in the NMR tube for more than 30s with only 1s LN$_2$ injection, during which period the temperature in the tube remains stable even with laser irradiation. Importantly, this implementation of sample freeze is completely compatible with the field cycling -- the LN$_2$ bath shuttled along with the sample at high speed and does not affect the inductive measurement at 7T. The sample can be rapidly thawed prior to measurement by a resistive heater attached in the NMR probe to enable high resolution readout.

\begin{figure*}[t]
  \centering
  {\includegraphics[width=0.96\textwidth]{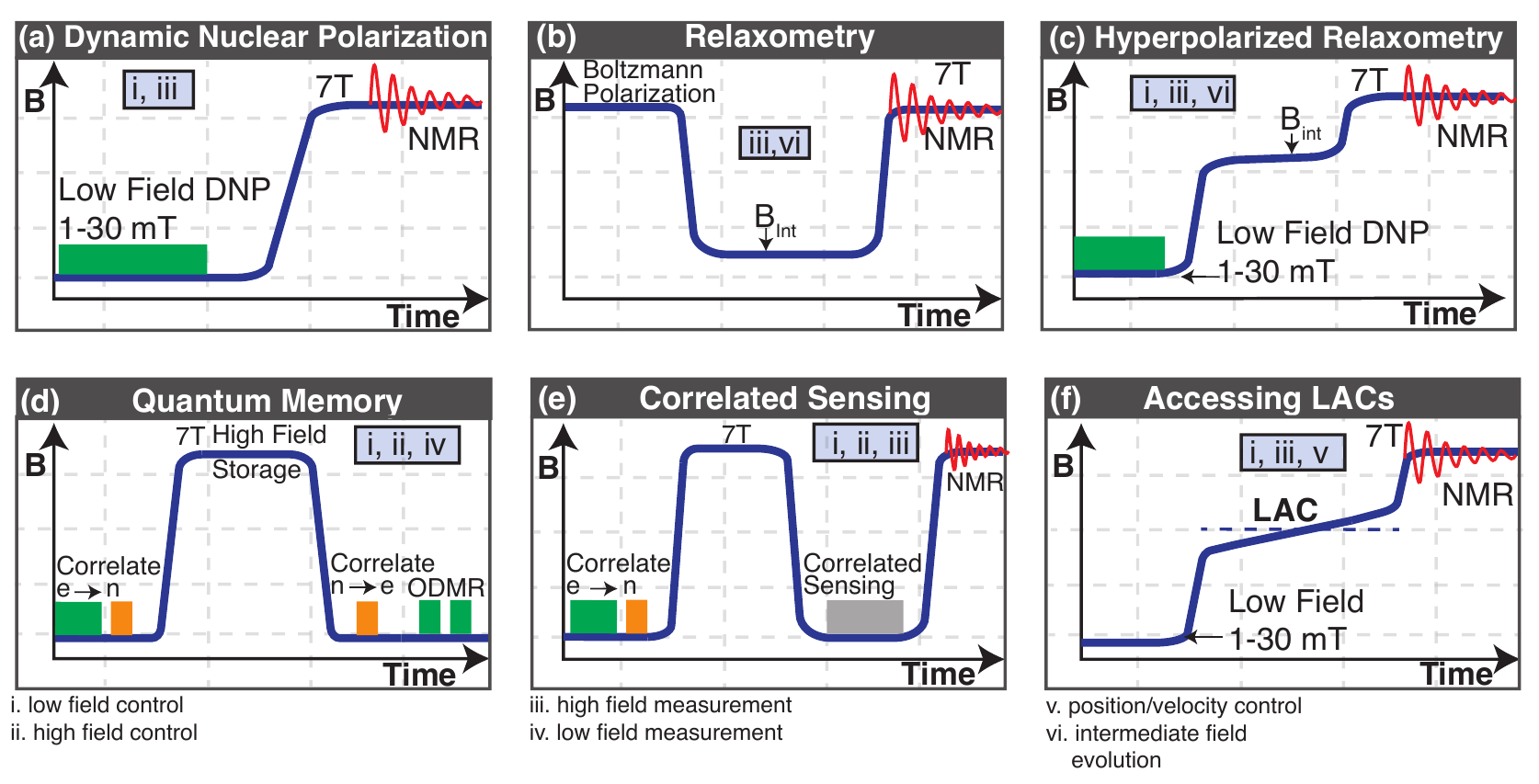}}
  \caption{\T{Proposed experiments} employing our field cycling instrument with low (1-30mT) and high (7T) field spin control and measurement (i-iv), along with the ability to access intermediate fields upto 2m/s sample shuttling speeds (v-vi). (a) \I{Low field DNP:} The device allows the room temperature hyperpolarization of $\Cs$ nuclei in diamond nano- and micro-particles, allowing their use in signal enhanced MRI. (b-c) \I{Relaxometry:} Accessing intermediate fields with high precision allows field dependent nuclear $T_1$ measurement with thermal polarization and hyperpolarization. (d-e) \I{Spin storage at high fields:} Long nuclear $T_1$'s at high field allow one to exploit the $\Cs$ nuclei as quantum memories for enhanced computation and sensing tasks. (f) The high spatial resolution allows the ability to access and sweep through level anti-crossings at intermediate fields with high precision, with applications in coherent polarization transfer.}
\zfl{toolbox}	
\end{figure*}
\section{Proposed experiments harnessing high and low magnetic fields}
\zsl{toolbox}

Our field cycling instrument allows one to harness the power of quantum control and measurement at low (1-30mT) and high (7T) fields, along with the ability to access intermediate fields with high resolution and controllable sweep rates. Indeed programming the field trajectories with a high degree of control allows versatile use of the device for several applications. In this section, we mention some potential experiments exploiting this capability. For concreteness, we focus  in \zfr{toolbox} on potential experiments in the system consisting of NV centers and $\Cs$ nuclear spins in diamond.

\I{Hyperpolarized diamond MRI:} -- The optical access and microwave control of the NV electronic spin available at low fields enable the coherent optical hyperpolarization of $\Cs$ nuclei (see \zsr{diamond}). Our instrument opens the potential to at once hyperpolarize diamond nano- and micro-particles at low field~\cite{Ajoy17}, and image them with MRI with high sensitivity (\zfr{toolbox}a). Diamond particles are non-cytotoxic, and can be easily functionalized~\cite{liu07,fu07,Schrand09,Bumb13}.  Hence the use of hyperpolarized $\Cs$ spins in diamond particles as MRI tracers is especially compelling -- the hyperpolarization providing bright MRI contrast that reports directly on the particle position, allowing their use as microfluidic flow tracers, as well as bio-sensors.

\I{Relaxometry:} -- Our device can be employed for relaxometry using conventional Boltzmann polarization (\zfr{toolbox}b) or low-field hyperpolarization (\zfr{toolbox}c).  Particularly pertinent for these applications are the ability to field cycle with high resolution, over a wide field range, and under programmable field trajectories, all of which our device can perform exceedingly well. We envision the use of this field cycling device broadly for hyperpolarized relaxometry of substances that are polarized via the diamond particles, with hyperpolarization providing significant savings in experimental time for the measurements. While in \zfr{T1B} we had considered exemplary hyperpolarized relaxometry on $\Cs$ nuclei in diamond, a more detailed exposition of the factors determining the $\Cs$ lifetimes (extracted from this data) will be presented in a forthcoming publication. 

\I{Quantum memories at high field:} --  High fields enable long $\Cs$ nuclear $T_1$'s, which in ultrapure diamond could exceed a few hours. Even for diamond microparticles containing a high density ($>$1ppm) of NV centers, we have observed nuclear $T_1$ approaching 400s at 7T (see \zfr{T1B}). These long lifetimes make the  $\Cs$ nuclei very attractive as ancillary quantum memories for the NV electronic qubits~\cite{Ajoy15,Zaiser16,Rosskopf17,Ajoy12g}. Field cycling hence allows the dual benefit of long nuclear lifetimes at high field and technically simple electronic control and readout at low field.  High fields are also associated with a concomitant increase in the nuclear coherence time, a fact that engenders their use in ancilla assisted quantum sensing protocols (\zfr{toolbox}e-f). Indeed there can be a significant boosts in quantum sensing resolution in such protocols, with resolution $\xD f\leq 1/T_{2,\R{memory}}$~\cite{Aslam17}. Such an experiment is schematically represented in \zfr{toolbox}e-f: a quantum sensing (eg. magnetometery) experiment is first performed via the NV electron, the states of the NV and $\Cs$ nuclei are then correlated~\cite{Laraoui13} and the quantum information stored at high field for $T_{2,\R{memory}}$. A subsequent sensing experiment performed with the NV can then be correlated with the previous measurement, and the result readout optically via NV center flouresence (\zfr{toolbox}e), or inductively via the $\Cs$ magnetization (\zfr{toolbox}f).

\I{Accessing level anti-crossings:} -- The high precision and repeatability of our instrument, along with its ability to sweep different fields with a programmable velocity profile, enable it to access specific level anti-crossings (LACs) over an extremely wide dynamic range (1mT-7T) (see \zfr{LAC}). LACs are traditionally helpful in several hyperpolarization contexts, for instance in optically pumped NV centers at the ESLAC ($\sim$ 510G) transferring their polarization to $\Cs$ nuclei and P1 centers~\cite{Fischer13,wunderlich17,Pagliero17}, and in parahydrogen based SABRE e.g. at 6mT to hyperpolarize ${}^{1}$H and at 0.1-1$\mu$T for $\Cs$ and ${}^{15}$N nuclei~\cite{Pravdivtsev213}. The high dynamic range allows access to low and high field LACs~\cite{Clevenson16} and clock states~\cite{Parker15x} that have been technically hard to probe previously. The small size of the sample and confined position at the center of the fringe field guarantee the homogeneity of the field.

\section{Conclusions and Outlook}
In summary, low and high fields offer complimentary advantages that can be harnessed together in order to open interesting avenues in metrology, hyperpolarization, and quantum information science. Low fields offer access to coupling dominated Hamiltonians, with low anisotropy, and spin indistinguishability; while high fields allow spin selectivity, long lifetime sand coherence times, and high SNR bulk detection. In this manuscript, we have described the construction and operation of a device that blends these two complimentary advantages on a single platform. In particular, we have constructed a novel field cycling platform capable of rapid ($<$ 700ms) magnetic field sweeps over a wide field range in principle up to 1nT-7T. High positional precision and repeatability allows access to fields with high resolution.  The device also allows optical and microwave spin control, and sample cooling at low fields, along with sensitive inductive readout at high field. While we have geared our device strongly to the applications in hyperpolarization of solids using optical pumping at low fields, we envision several applications in relaxometry, imaging, and quantum sensing, control and information storage that will be enabled by our instrument.

\section{Acknowledgements}

We gratefully thank A. Redfield, D. Suter, T. Zens and A. Llor for insightful conversations. We acknowledge valuable contributions from E. Granlund, J. D. Breen, S. Thangaratnam, R. Clark, S. Ebert, M. Garcia, D. Arnold, X. Cai, A. Kumar, A. Lin, G. Li, P. Raghavan, J. Wang, T. Huynh and J. Fyson, and the UC Berkeley College of Chemistry machine shop. C.A.M. acknowledges support from the National Science Foundation through grants NSF-1309640 and NSF-1401632, and from Research Corporation for Science Advancement through a FRED Award. The authors and UC Berkeley have filed patents on the instrument, techniques and applications.

\bibliography{C:/paper-drafts/Biblio}
\bibliographystyle{apsrev4-1}

\end{document}

%% file: Commands3.tex
\newcommand{\app}{\approx}

\newcommand{\Cs}{{}^{13}\R{C}}

\newcommand{\xD}{\Delta}

\newcommand{\mH}[0]{\mathcal{H}}

\newcommand{\beq}{\begin{equation}}
\newcommand{\eeq}{\end{equation}}
                  
\newcommand{\benum}{\begin{enumerate}}
\newcommand{\eenum}{\end{enumerate}}
                    
\newcommand{\bit}{\begin{itemize}}
\newcommand{\eit}{\end{itemize}}

\newcommand{\bea}{\begin{eqnarray}}
\newcommand{\eea}{\end{eqnarray}}

\newcommand{\bev}{\begin{verbatim}}
\newcommand{\eev}{\end{verbatim}}

\newcommand{\zt}{\times}

\newcommand{\T}[1]{\textbf{#1}}
\newcommand{\I}[1]{\textit{#1}}
\newcommand{\R}[1]{\textrm{#1}}
\newcommand{\zfl}[1]{\protect\label{fig:#1}}
\newcommand{\zfr}[1]{Fig. \ref{fig:#1}}

\newcommand{\zsl}[1]{\label{sec:#1}}
\newcommand{\zsr}[1]{Sec. \ref{sec:#1}}

\newcommand{\ba}{\left\{ \begin{array}{lr}}
\newcommand{\ea}{\end{array}\right.}

\newcommand{\blist}[1]{
 \begin{list}{#1}%$\ast\circ\bullet\Right
 \begin{align}
	 arrow
 \end{align}
 $\checkmark\star
  { \setlength{\itemsep}{3pt}
     \setlength{\parsep}{2pt}
     \setlength{\topsep}{3pt}
     \setlength{\partopsep}{0pt}
     \setlength{\leftmargin}{1em}
     \setlength{\labelwidth}{1em}
     \setlength{\labelsep}{0.5em} } }
\newcommand{\elist}{
  \end{list}  }

\DeclareMathSymbol{\vartheta}{\mathalpha}{letters}{"12}
\DeclareMathSymbol{\theta}{\mathalpha}{letters}{"23}
\DeclareMathSymbol{\phi}{\mathalpha}{letters}{"27}
\DeclareMathSymbol{\varphi}{\mathalpha}{letters}{"1E}

\newcommand{\bef}
{
\begin{figure}[htbp]
\centering
}

\newcommand{\eef}{\end{figure}}